\documentclass[aps,prl,reprint,showpacs,superscriptaddress]{revtex4-1}
\usepackage[dvipdfmx]{graphicx}

\usepackage{amsmath,amssymb}
\usepackage{color}

\usepackage{bm}
\usepackage{braket}
\usepackage{comment}

\newcommand{\figref}[1]{\figurename~\ref{#1}}

\begin{document}

%\preprint{}

\title{Einstein--de Haas fluctuation of a nanoparticle in spin polarized gases}



\author{Hideaki Nishikawa}
\affiliation{%
Kavli Institute for Theoretical Sciences, University of Chinese Academy of Sciences, Beijing, 100190, China.
}%
\affiliation{%
Department of Physics, Keio University, Hiyoshi, Kohoku-ku, Yokohama 223-8522, Japan
}%
\author{Daigo Oue}
\affiliation{%
Kavli Institute for Theoretical Sciences, University of Chinese Academy of Sciences, Beijing, 100190, China.
}%
\affiliation{%
The Blackett Laboratory, Department of Physics, Imperial College London, Prince Consort Road, Kensington, London SW7 2AZ, United Kingdom
}%
\author{Mamoru Matsuo}
\affiliation{%
Kavli Institute for Theoretical Sciences, University of Chinese Academy of Sciences, Beijing, 100190, China.
}%

\affiliation{%
CAS Center for Excellence in Topological Quantum Computation, University of Chinese Academy of Sciences, Beijing 100190, China
}%
%\email{}
\affiliation{
Advanced Science Research Center, Japan Atomic Energy Agency, Tokai, 319-1195, Japan
}
\affiliation{RIKEN Center for Emergent Matter Science (CEMS), Wako, Saitama 351-0198, Japan}

\date{\today}

\begin{abstract}
We theoretically study angular momentum (AM) transfer from a spin-polarized dilute gas into an nanoparitcle (NP) tightly trapped in optical tweezers. We formulate a microscopic model based on the spin tunneling Hamiltonian method and derive a macroscopic stochastic differential equation (SDE) which governs the AM-transfer-induced rotational motion of the NP. It is shown that the AM transfer rate at the NP surface can be extracted via the inference of the SDE. This work will open the door to the manipulation of nano-spintronic systems in gaseous environments.  
\end{abstract}

\maketitle %\maketitle must follow title, authors, abstract and \pacs

%%%%%%%%%%%%%%%%%%%%%%%%%%%%%%%%%%%%%%%%%%%%%%%%%%%%%%%%%%%%%%%%%%%%%%%
%%%%%%%%%%%%%%%%%%%%%%%%%%%%%%%%%%%%%%%%%%%%%%%%%%%%%%%%%%%%%%%%%%%%%%%
{\it Introduction.---}
The Einstein--de Haas (EdH) effect, mechanical rotation caused by spin polarization, is a universal phenomenon of angular momentum (AM) conversion in rigid bodies \cite{EdH1915}. The EdH effect has been demonstrated in a variety of magnetic systems \cite{Scott1962, Wallis2006, Ganzhorn2016, Dornes2019, Mori2020}.

Recently, the EdH effect has been utilized in micro-magneto-mechanical systems to measure spin relaxation processes involving electron spins \cite{Zolfagharkhani2008} and magnons \cite{Harii2019}.
These studies imply that the EdH effect will lead us to go further in studying the nonequilibrium spin physics in nano-scale materials if combined with a high-precision mechanical control and measurement.
Optical manipulation is a candidate for such mechanical control,
which was pioneered by Ashkin \cite{ashkin1970acceleration, ashkin1986observation}.
This technique has been applied to trap not only nanoparticles (NPs) \cite{gao2017optical} but also atoms, ions, and molecules \cite{chu1985three, bahns1996laser, vuletic2000laser}.
Furthermore, the optical angular momenta can be used as `spanners' to revolve tiny objects \cite{poynting1909wave, beth1936mechanical, simpson1997mechanical, padgett2011tweezers}. 
In the recent studies,
they further developed the optical trapping technique to cool the center-of-mass motion of NPs down to sub-Kelvin temperature \cite{chang2010cavity, gieseler2012subkelvin, millen2015cavity, tebbenjohanns2020motional}.
Those techniques will enable us to tightly trap and levitate NPs.
However, the magneto-mechanical response of such a levitated NP has not been studied so far.

\begin{figure}[htbp]
  \includegraphics[clip,width=8.0cm]{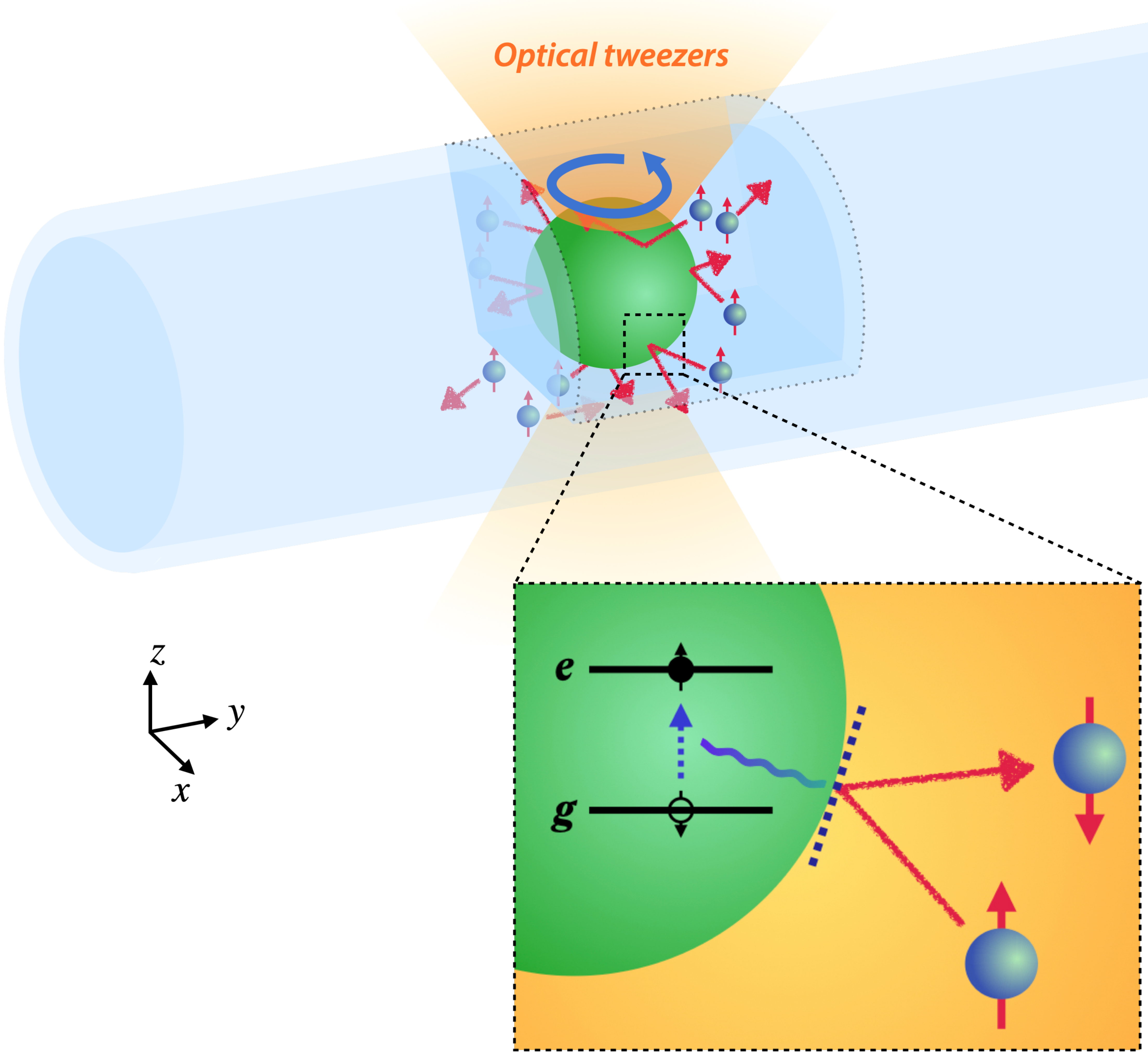}
  \caption{Setup of the Einstein--de Haas effect due to the spin transfer from the spin polarized gas into the NP trapped in a hollow-core optical fiber with optical tweezers. The spin AM of the gas is converted into the mechanical AM of the NP, whose electron state of the NP is modeled by a two-level quantum system consisting of down-spin (g) and up-spin (e) states as shown in the inset. The spins of the gas particles are injected at a certain rate only when there is a finite difference in the nonequilibrium spin state between the gas and the NP. The spin transfer rate can be determined by the stochastic differential equation for the rotational fluctuation of the NP.
  }
  \label{fig:fig1}
\end{figure}

In this Letter, we formulate the AM transfer from a spin-polarized dilute gas into a NP tightly trapped in an optical fiber with optical tweezers (\figref{fig:fig1}).
Our theory consists of (i) a microscopic model of the spin transfer and (ii) a macroscopic stochastic equation for the mechanical rotation of the NP.

The microscopic spin transfer from the gas to the NP is described by the spin tunneling Hamiltonian \cite{Ohnuma2014,Matsuo2018,Kato2019,Kato2020,Ominato2020a,Ominato2020b} where the spins can be injected from the gas into the NP only when there is a difference between the nonequilibrium distribution function of the gas and that of the NP,
and hence a spin chemical potential difference.

The macroscopic stochastic differential equation (SDE) is derived by applying the low noise approximation to the chemical master equation \cite{vanKampen2007}.
We reveal that the fluctuation of the NP rotation is described by the Ornstein--Uhlenbeck (OU) process. 
We also show that the spin transfer rate can be extracted from the variance of the OU process.

{\it Spin transfer from a spin-polarized gas into a NP.---}
Let us consider a microscopic description of the spin transfer from the spin-polarized gas into the NP trapped in the optical tweezers. 
The typical time scale of the spin transfer is considered to be much faster than the contact time of the gas at the surface of the NP. 
During the contact time, the spin of the gas is transferred at a certain rate into the NP. As a result, the spin polarization of the gas is flipped and the injected spin is converted into mechanical AM of the NP. 
We describe this situation by the following Hamiltonian:

\begin{align}
    H=H_\mathrm{np} + H_\mathrm{gas} + H_\mathrm{ex}, 
\end{align}
with a two-level Hamiltonian for the spin states in the NP $H_S$,
the Hamiltonian of the spin-polarized gas $H_B$,
\begin{align}
    H_\mathrm{np} &=\sum_{\sigma=\uparrow,\downarrow} \nu a_\sigma^\dagger a_\sigma, \quad
    H_\mathrm{gas} =\sum_{\boldsymbol{k}\sigma^\prime} \epsilon_{\boldsymbol{k}\sigma^\prime} c_{\boldsymbol{k}\sigma^\prime}^\dagger c_{\boldsymbol{k}\sigma^\prime},
\end{align}
and the spin exchange Hamiltonian between the NP and the gas,
\begin{align}
    H_\mathrm{ex} &=\sum_{\boldsymbol{q}} \lambda \sigma_{\boldsymbol{q}}^+ S^- + H.c.,
\end{align}
where $a_\sigma$ ($a_\sigma^\dagger$) and $c_{\boldsymbol{k}\sigma^\prime}$ ($c_{\boldsymbol{k}\sigma^\prime}^\dagger$) are the annihilation (creation) operators of the electron in the two-level system in the NP and in the gas, respectively.
We have introduced labels for the wavevector, $\boldsymbol{k}$, and for the spins, $\sigma$ and $\sigma^\prime$.
When the gas particles collide with the NP,
the spin AM is transferred from the gas into the NP with a tunneling amplitude of $\lambda$.
We have also defined spin flip operators,
\begin{align}
    S^- = a_{\downarrow}^\dagger a_{\uparrow},
    \quad
    \sigma_{\boldsymbol{q}}^+
        =\sum_{\boldsymbol{k}} c_{\boldsymbol{k}-\boldsymbol{q}\uparrow}^\dagger c_{\boldsymbol{k}\downarrow}.
\end{align}
The injected spins per unit time is given by the spin current at the NP surface defined as
\begin{align}
    &
    J_s := -\hbar \frac{\partial}{\partial t} \sigma_\textrm{tot}^z = i[\sigma_{\mathrm{tot}}^z, H],
    \\
    &
    \sigma_{\mathrm{tot}}^z := \frac{1}{2}\sum_{\boldsymbol{k}}                     
            (c_{\boldsymbol{k}\uparrow}^\dagger c_{\boldsymbol{k}\uparrow}
            - c_{\boldsymbol{k}\downarrow}^\dagger c_{\boldsymbol{k}\downarrow}).
\end{align}
Note that we assume that the collision between the spin-polarized and the NP is the perfect elastic collision and can safely ignore the contribution from the orbital AM.
Following the spin tunneling Hamiltonian method of the Schwinger-Keldysh formalism \cite{Matsuo2018,Kato2019},
we obtain the statistical average of the spin current:
\begin{align}
    \braket{J_s} 
    = \lambda^2 N_\mathrm{int} \int \frac{d(\hbar\omega)}{2\pi}\sum_{\boldsymbol{k}}
    \mathrm{Im} \chi^R_{\boldsymbol{k},\omega} (-\mathrm{Im} G^R_\omega)
    \delta f_\mathrm{ neq},
\end{align}
where $N_\mathrm{int}$ is the site number of two-level systems per unit area on the NP surface,
$\chi^R_{\boldsymbol{k},\omega}$ and $G^R_{\omega}$ are the Fourier components of the retarded spin susceptibility for the gas and of the retarded Green's function for the NP,
\begin{align}
    &
    \chi^R_{\boldsymbol{k},t}:=\frac{i}{\hbar}\theta_t \langle
    [\sigma_{\boldsymbol{k},t}^+, \sigma_{\boldsymbol{k},0}^-] \rangle,
    \quad
    G^R_t:=-\frac{i}{\hbar} \theta_t \langle [S^+_t, S^-_0] \rangle,
\end{align}
where $\theta_t$ is the unit step function.
The difference of the nonequilibrium distribution function $\delta f^\mathrm{neq}$ is given by
\begin{align}
    &\delta f^\mathrm{neq}_{\boldsymbol{k},\omega}
    =f^\mathrm{gas}_{\boldsymbol{k},\omega}-f^\mathrm{np}_\omega,
    \\
    &f^\mathrm{gas}_{\boldsymbol{k},\omega}
    :=\frac{\chi^<_{\boldsymbol{k},\omega}}{2i\operatorname{Im}\chi^R_{\boldsymbol{k},\omega}},
    \quad
    f^\mathrm{np}_\omega:=\frac{G^<_\omega}{2i\operatorname{Im}G^R_\omega},
\end{align}
where $\chi^<_{\boldsymbol{k},\omega}$ and $G^<_{\omega}$ are the Fourier transformations of the lesser components,
\begin{align}
    &
    \chi^<_{\boldsymbol{k},t} :=\frac{i}{\hbar}\langle
    \sigma_{\boldsymbol{k},0}^- \sigma_{\boldsymbol{k},t}^+ \rangle,
    \quad
    G^<_t:=-\frac{i}{\hbar}\langle S^-_0 S^+_t \rangle.
\end{align}

Note that the difference $\delta f^\mathrm{neq}$ should originate from the difference of the spin chemical potential between the gas and the NP,
that drives the spin transfer from the gas into the NP at a certain rate, which is proportional to $\lambda^2 N_\mathrm{int}$. 
We will show how the spin transfer rate can be extracted from a fluctuating rotational motion of the NP in a possible situation below. 

{\it EdH effect of the NP.---}
The injected spins from the gas are converted into the mechanical AM of the NP via the EdH effect. 
The AM conservation among the gas and the NP is given by
\begin{align}
    \braket{\sigma^z_\mathrm{tot}(t)} + I \omega(t) = \mathrm{const}., 
\end{align}
where $\sigma_\mathrm{tot}^z(t)$ is the total amount of the spin accumulation in the  spin-polarized gas defined as
\begin{align}
    \braket{\sigma^z_\mathrm{tot}(t)}:= \braket{\sigma^z_\mathrm{tot}(0)}-\int^{t}_{0} dt^\prime \braket{J_s(t^\prime)}.
\end{align}
The NP moment of inertia is $I$, and the angular velocity of the NP is $\omega(t)$.

{\it SDE of the EdH fluctuation. ---}
In order to connect the microscopic description of the spin transfer processes to the macroscopic quantities, we construct a SDE governing the EdH-induced rotational motion of the NP. 
For simplicity, we assume that the injected spins are only converted into the mechanical AM parallel to the spins, and the NP is perfect sphere so that the collision between the gas and the NP is elastic. 

In a dilute gas, the spin transfer in each collision is independent,
and thus we can describe it by a counting process. 
This is where the stochasticity arises in our system.
The counting process can be regarded as a pair of chemical reactions in the system with a volume of $\Omega$,
\begin{align}
    \begin{cases}
    A_1 \overset{\kappa_1}\longrightarrow A_2
    &\textrm{(Reaction 1)},
    \\
    A_2 \overset{\kappa_2}\longrightarrow A_1
    &\textrm{(Reaction 2)},
    \end{cases}
    \label{eq:reaction}
\end{align}
where $A_1$ and $A_2$ are a spin-polarized (up-spin) gas particle and non-polarized (down-spin) one.
We have defined $N_{1,2}$ as the number of up-spin particles and that of down-spin particles. 
Note that the total particle number $N:=N_1+N_2$ is conserved.
The reaction 1 represents the spin transfer from a gas particle into the NP while the reaction 2 corresponds to its back-action from the NP to the gas particle. 
The reaction rate per unit time,
$\kappa_m (m=1,2)$,
is associated with the spin current via a micro-macro correspondence, 
\begin{align}
    \braket{J_s}=\kappa_1N_1-\kappa_2N_2.
    \label{eq:correspondence}
\end{align}

The stochastic dynamics of the particle numbers,
$\vec{N}:=(N_1,N_2)^{\top}$,
is subject to the chemical master equation (CME),
\begin{align}
    \frac{\partial P_t(\vec{N})}{\partial t} 
    = \sum^{2}_{m=1}
    \Big[&F_m(\vec{N}-\vec{s_m})P_t(\vec{N}-\vec{s_m})%\right.
    \notag \\
    %&\hspace{3.5em}
    &-F_m(\vec{N})P_t(\vec{N})\Big],
    \label{eq:CME}
\end{align}
where $P_t(\vec{N})$ is the probability distribution function of $\vec{N}$,
and $F_m(\vec{N}):=\kappa_m N_m$ is the propensity functions of the reaction $m$.
The stoichiometric vectors accompanying the reaction $m$,
\begin{align}
    \vec{s}_{m=1}:=
        \begin{pmatrix}
        -1 \\ +1
        \end{pmatrix},
    \quad
    \vec{s}_{m=2}:=
        \begin{pmatrix}
        +1 \\ -1
        \end{pmatrix},
\end{align}
are responsible for the variation of the particle numbers of $A_1$ and $A_2$.

If the stoichiometric change is small compared to the total reactants population,
$|\vec{s_m}| \ll |\vec{N}|$,
the CME can be approximated by a continuous diffusion process. 
Here,
in order to derive the SDE governing the NPAM dynamics,
we use the system size expansion,
that is also called as the low noise approximation and valid within large system and small fluctuation \cite{vanKampen2007}.
Firstly, we separate the particle numbers in terms of deterministic and fluctuating parts of the number densities,
\begin{align}
    \vec{N} = \Omega \vec{\phi}(t) + \Omega^{1/2} \vec{\xi}(t). 
    \label{eq:phi_xi}
\end{align}
It is worth noting that it is generally important to analyze the fluctuating part $\vec{\xi}(t)$ in small systems including the NP trapped by the optical tweezers.
The deterministic part $\vec{\phi}(t)$ is the solution of the chemical rate equations,
\begin{align}
\frac{\partial}{\partial t} \vec{\phi}(t)= K \vec{\phi}(t),\quad
K=\begin{pmatrix}
-\kappa_1 & \kappa_2 \\
\kappa_1 & -\kappa_2 \\
\end{pmatrix}.
\end{align}

Within the small stoichiometric change regime
(i.e., $|\vec{s_m}| \ll |\vec{N}|$),
we can perform the Kramers-Moyal expansion around $\vec{N}$ for $\vec{s_m}$. 
\begin{align}
    \frac{\partial P_t(\vec{N})}{\partial t}
    &= \sum_{\substack{l_1,l_2 \geq 0 \\ m=1,2}} 
    \Gamma_{l_1,l_2}^m
    \frac{\partial^{l_1+l_2}}{{\partial N_1}^{l_1}{\partial N_2}^{l_2}}
    \left[F_m(\vec{N})P_t(\vec{N})\right]
\end{align}
where we have defined
$\Gamma_{l_1,l_2}^m \equiv (-s_{1,m})^{l_1} (-s_{2,m})^{l_2}/l_1!l_2!$.
Substituting Eq. \eqref{eq:phi_xi},
we write the probability distribution function as a function of re-scaled variable $\vec{\xi}$ instead of $\vec{N}$,
\begin{align}
    P_t(\vec{N})=\Pi_t(\vec{\xi}),
\end{align}
and collect the zeroth order terms in the system size $\Omega$ to get the Fokker-Planck-type diffusion equation for $\vec{\xi}$.
\begin{align}
    \frac{\partial \Pi_t(\vec{\xi})}{\partial t}
    &= \sum_{\substack{l_1,l_2 \geq 0 \\ l_1+l_2= 1\\ m = 1,2}}
        \Gamma_{l_1,l_2}^m
        \frac{\partial}{{\partial \xi_1}^{l_1}{\partial \xi_2}^{l_2}}
            [\kappa_m \xi_m \Pi_t(\vec{\xi})] \notag \\
    &+ \sum_{\substack{l_1,l_2 \geq 0 \\ l_1+l_2=2 \\ m = 1,2}}
        \Gamma_{l_1,l_2}^m
        \frac{\partial^2}{{\partial \xi_1}^{l_1}{\partial \xi_2}^{l_2}}
        [\kappa_m \phi_m \Pi_t(\vec{\xi})]
\end{align}
We can get the corresponding Langevin-type SDE \cite{vanKampen2007},
\begin{align}
    d \xi_i(t) 
    &= \sum^{2}_{m=1} \Big[ s_{i,m} \kappa_m \xi_m(t) dt 
    \notag \\
    &+ s_{i,m} \sqrt{\kappa_m \phi_m(t)} \cdot d W_m (t) \Big]
    \quad (i=1,2).
\end{align}

In the following,
we consider a situation where almost all of the gas particles are spin-polarized [i.e., $N_1 (0)\approx N, N_2 (0)\approx 0$]
in the initial state and focus on the short-time dynamics so that we can safely assume that $N_1\gg N_2$.
From these observation, we can assume that the first component of $\vec{\xi}$ is much larger than the second one, $|\xi_1| \gg |\xi_2|$.
Since we focus on the short-time dynamics after the initial state,
we have $\phi_i (t) \sim \phi_i (0)=\mathrm{const}.$ and $\phi_1 (t) \gg \phi_2 (t)$.
In other words, 
we can ignore the back-action from the NP to the gas particles. Finally, we can write the SDE of the spin AM
\begin{align}
    d \xi_1 (t) \approx -\kappa_1\xi_1 dt + \sqrt{\kappa_1\phi_1 (0)} \cdot dW_1 (t).
    \label{eq:LE_approx}
\end{align}
This equation describes the fluctuations of spin AM of NP.
This type of SDEs is often called as the OU process \cite{OUP1930}.

We perform the parametric estimation of this stochastic process to estimate the reaction rate $\kappa_1$ and the spin current $\langle J_s \rangle$ by using the micro-macro correspondence \eqref{eq:correspondence}. 
The mean and variance of the OU process are evaluated as follows:
\begin{align}
    &
    \langle \xi_1 (t) \rangle = \xi_1 (0) e^{-\kappa_1 t},
    \quad 
    \operatorname{var}[\xi_1 (t)] = \frac{N}{2\Omega}(1-e^{-2\kappa_1 t}).
    \label{eq:mean_variance}
\end{align}
The relation between the angular velocity of the NP $\omega(t)$ and the fluctuation $\xi_1(t)$ is given by means of the AM conservation,
\begin{align}
    I\frac{d\omega(t)}{dt}=\kappa_1(\Omega\phi_1(t)+\Omega^{1/2}\xi_1(t)).
\end{align}

\begin{figure}[htbp]
  \includegraphics[clip,width=.9\linewidth]{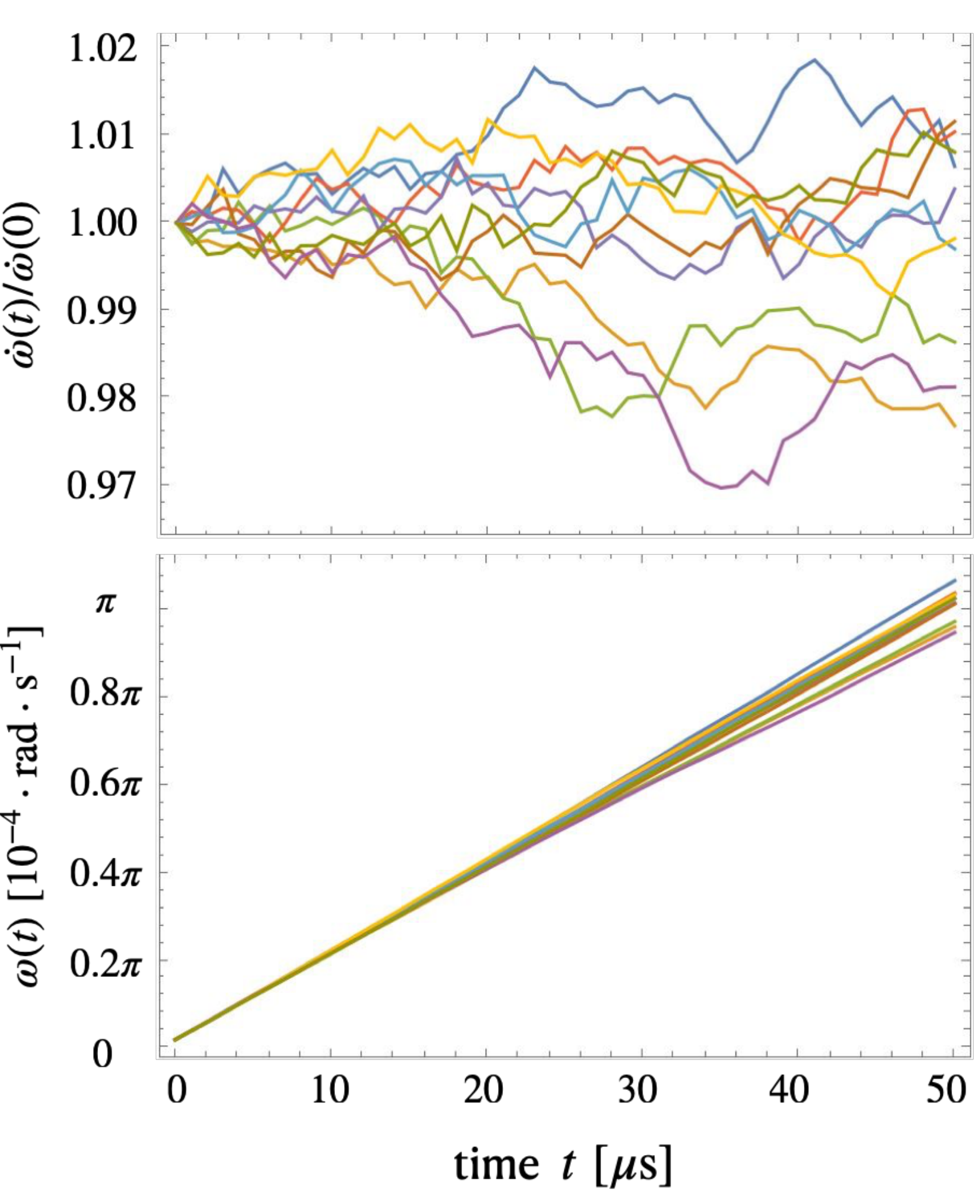}
  \caption{Numerical solutions of the SDE (Eq.\eqref{eq:LE_approx}) with $\kappa_1=1$, $N=100$, $\Omega=1$, and $\xi_1(0)=0$.
  (a) The time evolution trajectories of the NP angular acceleration $\Dot{\omega}(t)$ and (b) that of the NP angular velocity $\omega(t)$ for different pseudo-random numbers.
  }
  \label{fig:mean_variance}
\end{figure}

{\it Discussion.---}
The trajectory data generated by Eq. \eqref{eq:LE_approx} is shown in \figref{fig:mean_variance}.
We can calculate the mean and the variance in accordance with Eq. \eqref{eq:mean_variance} to estimate the reaction rate $\kappa_1$.
We can also evaluate the spin current $\braket{J_s}$ by the micro-macro correspondence \eqref{eq:correspondence}.
We have estimated the angular acceleration of the NP as follows:
\begin{align}
    \frac{d\omega}{dt} \sim \tau^{-1} N \hbar \frac{V}{\Omega}I^{-1} \sim 10^{1}\ \mathrm{[rad \cdot s^{-2}]},
\end{align}
where $\tau$ is the time interval of the collision per one spin-polarized gas, which is evaluated as $\tau \sim R/\sqrt{k_\mathrm{B}T} \sim 0.1 \ \mathrm{[\mu s]}$ where $R\sim 10\ \mathrm{[\mu m]}$ is the core radius of the optical fiber, and $k_\mathrm{B}$ is the Boltzmann's constant. 
The number of the gas particles is estimated $N\sim10^{5}$ at a pressure of $10^{-3}\  \mathrm{[Pa]}$.
The volume of the NP and that of the optical fiber are estimated at $V\sim r^3 \sim 10^{-21}\ \mathrm{[m^3]}$ and $\Omega \sim R^3 \sim 10^{-15}\mathrm{[m^3]}$.
The NP moment of inertia $I$ is in the order of $\sim 10^{-31}\ \mathrm{[J \cdot m^2]}$. Note that the transition rate from a gas particle to the NP is assumed to be $100 $\%.

Finally, let us mention the contribution from the orbital AM of the gas to the NP rotational motion. We have assumed the perfectly spherical NP so that the collisions between the NP and the gas particles are elastic. Thus, we can safely neglect  peripheral collisions of the gas and hence the contribution from the orbital AM. 
When the collision is inelastic, although both the spin AM and the orbital AM drive the NP rotational motion, we can separate them because the orbital AM only contributes to the variance of the angular acceleration while the spin AM contributes to both the mean and the variance.

{\it Conclusion.---}%
In this Letter, we have investigated the AM transfer between a spin-polarized gas and a NP trapped by optical tweezers. To describe the EdH fluctuation caused by the spin transfer from the gas into the NP, we formulated a microscopic model of the spin transfer and a macroscopic SDE for the fluctuating rotational motion of the NP. 
Using the spin tunneling Hamiltonian method, we modeled the microscopic process of the spin transfer from the gas into the NP, whose spin state is described by a two-level quantum system. 
We showed that the spins are injected at a certain rate when there exists a difference in the nonequilibrium spin chemical potential between the gas and the NP. 
We performed the low noise approximation of the CME to derive a macroscopic SDE of OU type for the fluctuation of the NP angular velocity and acceleration. 
We found that the spin transfer rate can be extracted from the variance of the OU process. 
The present results will offer a tool for investigating nonequilibrium spin states at the nanoscale, combined with high-precision measurement techniques such as optical tweezers.

{\it Acknowledgements.---}
The authors are grateful to K. Saito and H. Chudo for valuable comments.
This work is partially supported by the Priority Program of Chinese Academy of Sciences, Grant No. XDB28000000 and 
Grant-in-Aid for Scientific Research B (20H01863) %mamoru kiban-B
from MEXT, Japan.

\bibliography{ref}

\end{document}